\documentclass[11pt]{article}
\usepackage{times, subfigure}
\usepackage{alltt}
\IfFileExists{upquote.sty}{\usepackage{upquote}}{}
\usepackage{framed}
\usepackage{color}
\usepackage{harvard}
\usepackage{url}
\usepackage{epsfig,setspace,rotating,fullpage}
\def\eqx"#1"{{\label{#1}}}
\def\eqn"#1"{{\ref{#1}}}

\definecolor{fgcolor}{rgb}{0.345, 0.345, 0.345}
\newcommand{\hlnum}[1]{\textcolor[rgb]{0.686,0.059,0.569}{#1}}%
\newcommand{\hlopt}[1]{\textcolor[rgb]{0,0,0}{#1}}%
\newcommand{\hlstd}[1]{\textcolor[rgb]{0.345,0.345,0.345}{#1}}%
\newcommand{\hlkwb}[1]{\textcolor[rgb]{0.69,0.353,0.396}{#1}}%
\newcommand{\hlkwd}[1]{\textcolor[rgb]{0.737,0.353,0.396}{\textbf{#1}}}%

\usepackage{framed}
\makeatletter
\newenvironment{kframe}{%
 \def\at@end@of@kframe{}%
 \ifinner\ifhmode%
  \def\at@end@of@kframe{\end{minipage}}%
  \begin{minipage}{\columnwidth}%
 \fi\fi%
 \def\FrameCommand##1{\hskip\@totalleftmargin \hskip-\fboxsep
 \colorbox{shadecolor}{##1}\hskip-\fboxsep
     \hskip-\linewidth \hskip-\@totalleftmargin \hskip\columnwidth}%
 \MakeFramed {\advance\hsize-\width
   \@totalleftmargin\z@ \linewidth\hsize
   \@setminipage}}%
 {\par\unskip\endMakeFramed%
 \at@end@of@kframe}
\makeatother

\definecolor{shadecolor}{rgb}{.97, .97, .97}
\definecolor{messagecolor}{rgb}{0, 0, 0}
\definecolor{warningcolor}{rgb}{1, 0, 1}
\definecolor{errorcolor}{rgb}{1, 0, 0}
\newenvironment{knitrout}{}{} 

\usepackage{color}
\definecolor{gray}{rgb}{0.5,0.5,0.5}

\title{
Challenges and opportunities for statistics and statistical education: \\ looking back, looking forward}
\author{Nicholas J. Horton\thanks{
Address for correspondence: Department of Mathematics and Statistics, Amherst College, AC\#2239, PO Box 5000, 
Amherst, MA  01002-5000.  Phone: 413-542-5655, email:
nhorton@amherst.edu}
\\
\footnotesize 
Department of Mathematics and Statistics\\
\footnotesize 
Amherst College, Amherst, MA
\normalsize
}
\begin{document}
\maketitle
\newpage
\begin{center}
{\large
Challenges and opportunities for statistics and statistical education: \\ looking forward, looking back}
\end{center}

\subsection*{Abstract}

The 175th anniversary of the ASA provides an opportunity to look back into the past and peer into the future.  
What led our forebears to found the association?  What commonalities do we still see?  What insights
might we glean from their experiences and observations?  I will use the anniversary
as a chance to reflect on where we are now and where we are headed in terms of statistical education
amidst the growth of data science.  Statistics is the science of learning from data.  By fostering more multivariable thinking, building data-related skills,
and developing simulation-based problem solving, we can help to ensure that 
statisticians are fully engaged in data science and the analysis of the abundance of data now available to us.  

Keywords: 
American Statistical Association,
data science,
empirical problem solving, 
history,
Lemuel Shattuck,
simulation studies,
statistical computing, 
statistical education

\section{Looking back}

The 175th Anniversary of the American Statistical Association (ASA) has special meaning in the City of Boston, Massachusetts.  Besides being home to the Boston Chapter of the ASA (proudly serving members in Massachusetts, Vermont, New Hampshire, Maine, and Rhode Island), the association was founded there on November 27, 1839 and has been active ever since.  Any anniversary provides an opportunity to look back as well as forward, and this celebration is no exception.  

A knowledge of history helps to ground our understanding and provides insights that might be useful for the future.
I'd like to start by discussing the contributions of
Lemuel Shattuck (1793--1859), one of the five original founders of the ASA \cite{will:1940,will:1947}.  Along with William Cogwell (former pastor and agent for the American Education Society), Richard Fletcher (lawyer), John Dix Fisher (physician), and Oliver Peabody (lawyer, clergyman, and poet), Shattuck worked to improve the quality and use of statistics in Boston, Massachusetts, and beyond.  According to our current standard, the founders would not be considered statisticians (not surprisingly, since the profession did not exist).  
Raymond \citeasnoun{pear:1940} described this somewhat peculiar group as:
\begin{quote}
an odd lot of fish, differing widely from each other in most respects, but all alike in one.
Each of them had what the psychiatrists nowadays call a compulsion neurosis impelling him to tinker
with numbers and fiddle with figures.  Their souls cried out for tabulations in the same
way that the prohibitionist of later times yearned for his daily ration of Peruna.
\end{quote}

While much of the initial work of the ASA and its members involved somewhat mundane but still important efforts to make registration of births, marriages, and deaths more effective, Shattuck helped to pioneer American public health through his work on a sanitary survey of Massachusetts \cite{shat:1850,shat:1959}.
Using registries, graveyard records, and church records, 
Shattuck identified important health disparities.

\begin{figure}[tbh]
\begin{center}
\includegraphics[height=3.8in,width=4.8in]{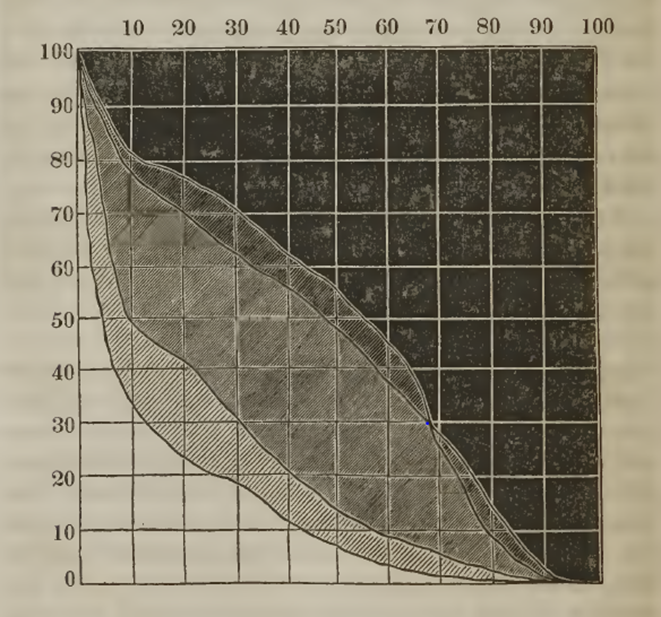}
\caption{Estimated survival for various populations in the early 19th century: top line represents Preston England (``healthy classes in England"); second from top residents of Newton, Massachusetts; third from top residents of Boston, Massachusetts; bottom Boston Catholics. Dramatic disparities in longevity are seen, particularly in the first ten years of life.}
\label{fig:shattuck}
\end{center}
\end{figure}

The only graphical image in the report provides a stark depiction of a dramatic health gradient.
Figure \ref{fig:shattuck} displays
the estimated survival (Y-axis) as a function of age (X-axis) of a hypothetical cohort from one of four populations (from top to bottom): Preston, England (``healthy classes in England"), those living in Newton (a community next to Boston that consisted primarily of farmland), residents of Boston, and Catholic residents in Boston.  The differences are striking: only 34\% of Boston Catholics survived to age 10, while
nearly 80\% of those in neighboring Newton survived to that age.  Shattuck alternated
between understatement:
(``But persons die at all ages, and in some classes very much earlier than in others.") and righteous fervor:
\begin{quote}
{\it It is proved} (Shattuck's italics) that causes exist in Massachusetts, as in England, to produce premature and preventable deaths, and hence unnecessary and preventable sickness; and that these causes are active in all the agricultural towns, but press most heavily upon cities and populous villages.
\end{quote}
While it took many years for Shattuck's advice to be heard 
(and we still face many important health disparities), 
this is by any account a remarkable display of the staggering disparities in health outcomes.  
When we think about ways to communicate the results of a statistical study, Shattuck's
graphical display is still instructive.  In addition, though it predates the development of modern methodology, the nonparametric nature harkens to the 
culture of algorithmic modeling described by \citeasnoun{brei:2001}.

What about the experiences of those who celebrated the centennial of the ASA founding?
The leaders of the ASA in 1939 had benefit of a century of hindsight and a far larger and more sophisticated 
association. They faced many issues eerily familiar to those we observe now, noting how rare it was for 
statistical methods and approaches to be used and promulgated by 
\emph{professional statisticians}:
``Yet even today, professional statisticians are few among the many who, more or less expertly,
use statistical tools and materials in diverse professions and occupations"  \cite{davi:1940}.  Then as now statistics are often used by many individuals who are not formally trained in the discipline. It remains critically important that these methods and techniques are appropriately used.

But even these leaders in the field found predictions challenging.  The 100th anniversary celebrations estimated the peak total population of 
North America to be between 160 to 180 million \cite{davi:1940}, well below the United States population estimate of 318,857,056 by
the Census Bureau as of July 1, 2014 (see \url{https://www.census.gov/popest/about/index.html}).  While they should be commended for their use of an interval for the estimated peak population, 
their estimates widely missed the mark.

\section{Where are we now?}

The experience of our professional forebears proves
the truism that prediction is challenging.  But the 175th anniversary of the founding of the ASA tempts one 
to indulge in the practice.  
Turning to the present and the future, where do we stand as an association and a profession?
What are our internal strengths and weaknesses?  What are the external threats and opportunities
before us?
How do we ensure that when we look back at our 200th anniversary that statistics will continue to be a thriving discipline as well as a vibrant choice for our students?

I will start with some encouraging developments.  Interest in the discipline of statistics and the analysis of
data
is booming.  George Lee of Goldman Sachs estimates that 90\% of the world's data have been created in the last two years (\url{http://www.goldmansachs.com/our-thinking/trends-in-our-business}).  These increasingly diverse data are being used to make decisions in all realms of society.  As but one example, consider the theme for the AAAS annual meeting in 2015 (Innovations, Information, and Imaging): ``Science and technology are being transformed by new ways to collect and use information. Progress in all fields is increasingly driven by the ability to organize, visualize, and analyze data" (\url{http://meetings.aaas.org/program/meeting-theme}).  
The widely cited McKinsey report (\url{tinyurl.com/mckinsey-nextfrontier}) described the potential shortage of hundreds of 
thousands of workers with the skills to make sense of the enormous amount of information now available.

\begin{figure}[tbh]
\begin{center}
\includegraphics[height=3.8in,width=4.8in]{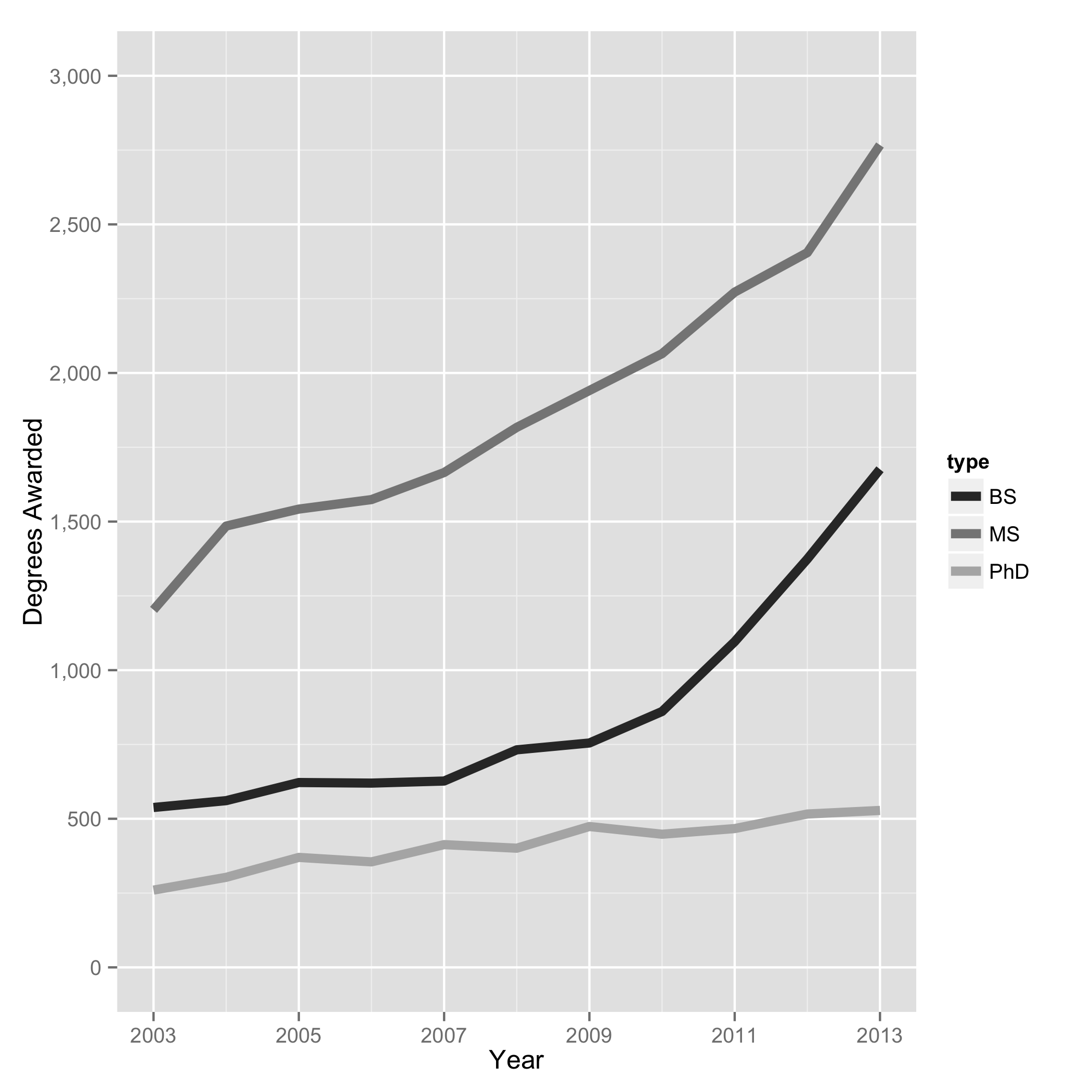}
\caption{Graduation numbers over time for statistics programs: master's degrees (top line), bachelor's degrees (middle line), and doctoral degrees (bottom line).  Source: IPEDS (Integrated Post-secondary Education Data System Completions Survey)}
\label{fig:undergrad}
\end{center}
\end{figure}

Encouraging growth is being seen in statistics degree programs, particularly at the master's and bachelor's level.
Figure \ref{fig:undergrad} displays the number of students completing master's (top line),  
bachelor's (middle line), and doctoral degrees (bottom line) in the United States through the year 2013.
Unlike most fields (such as psychology or mathematics), where the number of bachelor's graduates far outnumbers the number of master's graduates, statistics has more graduates at the master's level.  This is likely a historical artifact, since an undergraduate degree in statistics is a relatively recent development.  

The growth of undergraduate programs in statistics is an important development to help meet societal demand.  
The emergence of statistics as a distinct discipline, not an add on to mathematics for highly educated specialists, is relatively new.
Recent
guidelines for undergraduate programs in statistics provide a framework to ensure that graduates have the necessary skills to make contributions from day one in the workforce \cite{asa:2014}.  Being able to succeed at this level of training is critical, given the cost of higher education: it may not be financially feasible for all students to complete a four-year undergraduate degree and then pursue an additional one- or two-year master's program before being productive.

Where will the hundreds of thousands of new workers anticipated by the McKinsey report and others come from?
Graduates of statistics programs will be a small fraction (even if the growth seen in the recent decade continues or
accelerates).  
It's likely not to be solved by an influx of new statistics doctoral students.  
While the number of doctoral graduates is slowly increasing, growth is insufficient to meet demand for new positions in industry, government, and academia.  

Where else can these skilled graduates be found?  If we don't produce them, who will?
The 2013 Future of Statistics (London) report (\url{http://bit.ly/londonreport}) describes the 
need for ``data scientists"---the exact definition of which is still a matter of
debate---and raises important questions about the identity and role of statisticians.  What training is needed to be able to function in these new positions?  What role does statistics have in this new arena?  How do we ensure that those not formally trained in advanced statistics have sufficient appreciation and background?  These are similar to questions that the founders of the ASA and those who peered backwards and 
forwards at the 100th anniversary considered \cite{davi:1940}.

A widely read Computing Research Association white paper on the challenges and
opportunities with `Big Data' starts in a familiar manner:
``The promise of data-driven decision-making is now being recognized broadly, and there is growing enthusiasm for the notion of `Big Data'" (\url{http://www.cra.org/ccc/files/docs/init/bigdatawhitepaper.pdf}).  But it is disconcerting that the first mention of statistics is not found until the sixth page of the report: ``Methods for querying and mining Big Data are fundamentally different from traditional statistical analysis on small samples."  The remaining references include statistics in passing as a bag of tricks (but not central to the use of data to inform decision-making).  The London report warned that unless statisticians engage in related areas that are perhaps less familiar there is a potential for the discipline to miss out on the important
scientific developments of the 21st century.  

\begin{figure}[tbh]
\begin{center}
\includegraphics[height=3.8in,width=4.8in]{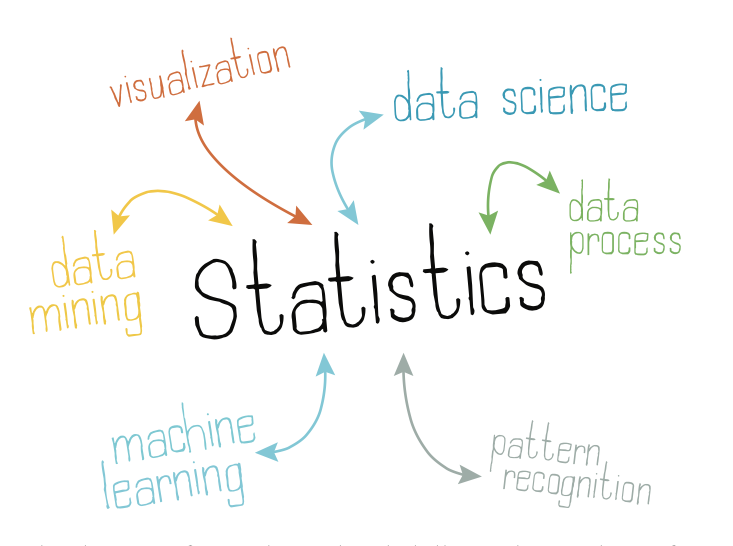}
\caption{One view for the ways that statistics (the science of learning from data) integrates with other key topics. Source: American Statistical Association}
\label{fig:datascience}
\end{center}
\end{figure}

In my work as an applied biostatistician, I've seen firsthand the importance of statistics to ensure that scientific investigations are on a solid foundation.  
I agree with the definition that 
statistics is the science of learning from data \cite{vdlaan:2015}.
The appropriate use of statistics ensures that variability and bias are addressed, suitable analytic methods are undertaken, interpretations are rational and defensible, and decisions made to account for uncertainty. 
We need to ensure that statisticians are firmly embedded in data science. 
Figure \ref{fig:datascience} provides one schematic for what this might involve.

The bidirectional arrows are intentional: there are strong connections between statistics and these related areas such as visualization, machine learning, and data science, though it's not appropriate to say that all are wholly subsumed by statistics.  Many of these areas are familiar to most of us, at least through publications such as \citeasnoun{nola:temp:2010}, while some may be outside our traditional training (and hence, our comfort zone).  Anecdotal reports have indicated that statistics undergraduates are at a competitive disadvantage relative to undergraduate computer science (CS) degree holders for entry-level positions.  
Such positions tend to have data-related skills at the core of the job descriptions. At present, many CS students tend to be able to perform computations with data more easily than their statistics equivalents.  
This shouldn't be allowed to continue.  
How do we ensure that our students (and those in other quantitative disciplines) are able to effectively make sense of the data around them?  How do we fully engage with all of the topics included in Figure \ref{fig:datascience}?

\section{Looking forward: what and how we teach}

In my role as an instructor, I've seen the challenges and opportunities in training the next generation.  How can we provide more opportunities for students (not just the limited numbers who end up majoring in statistics) to ``think with data" as Diane Lambert of Google has so eloquently described?  How do we open up our field as \citeasnoun{brow:kass:2009} proposed?  We need to start by considering our first courses.  All too often students emerge from a first or second course in statistics with the perception that statistics involves memorization of the use of ``cookbook" methods and the postulation and rote application of null hypothesis statistical procedures.  Some of the examples that we have 
historically used (e.g., the colors of M\&Ms) do not convey the potential for our methods in more compelling settings \cite{goul:2010}.  
Such activities---while often popular with students---don't really mirror the use of statistics in the real world.

While the \citeasnoun{gaise} report encouraged the use of technology (which is widespread in most courses), hundreds of thousands of high school students still use outmoded calculators for their analysis, limiting their ability to move beyond simple calculations or undertake any sense of realistic workflow that they might encounter in the real world.  This is certainly not the technology being used by data scientists.

As we ponder how to adjust how and what we teach, I propose three ways to address these challenges: 
(1) broaden the role of multivariable methods in our curricula; (2) develop data-related skills early; and (3) expand the role of simulation and computation.  In the next sections I will argue for why these are important and indicate how they might be incorporated into our courses and programs.

\subsection{Multivariate thinking and the basics of confounding and causal inference}

Issues of confounding and bias arise commonly in many analyses that are labeled as data science.  Analysts working in these areas must be able to understand issues of design, confounding, and bias \cite{kaplan:fresh}.  
This is an area where statisticians bring great value.

The lack of appreciation for simple multivariable methods is a major limitation in too many of our courses.
Students are generally taught that 
if data arise from well conducted randomized trials, they can make causal conclusions using a two-sample t-test.  
All too often, this is considered the pinnacle of statistics (see \citeasnoun{cobb:2007} for a dissenting view).  
But most data that students see are not derived from a randomized trial with no dropout, full adherence, and sufficient blinding.  In these situations, students are stymied by the bivariate coverage of topics in the syllabi of the Advanced Placement Statistics and most 
intro stats courses.  I worry that students may be paralyzed by what is likely their only statistics course \cite{meng:2011}
and not see the full potential for statistics as a foundation 
for learning from data.

The new ASA guidelines for undergraduate programs in statistics state that 
students need a clear understanding of principles of statistical design
and tools to assess and account for the possible impact of
other
measured and unmeasured variables \cite{asa:2014}.
This can't all happen in a single statistics course, but it is important that students are exposed to the basic principles
early and often.

\begin{figure}[tbh]
\begin{center}
\includegraphics[height=3.8in,width=4.8in]{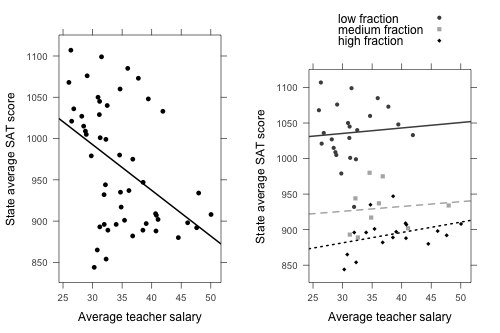}
\caption{Left: unconditional association of state average teacher salary with average SAT score; Right: association after accounting for the fraction of students taking the SAT in that state}
\label{fig:sat}
\end{center}
\end{figure}

What can we teach students in the first course?  Consider an example where statewide data from the mid-1990's are used to assess the association between average teacher salary in the state and average SAT (Scholastic Aptitude Test) scores for students \cite{gube:1999}.  These exams are used for college entry, and the results are sometimes used as a proxy for educational quality.  The leftmost
graph in Figure \ref{fig:sat} displays the unconditional association between these variables.  There is a statistically significant negative relationship. The model predicts that a state with an average salary that is one thousand dollars 
higher than another would have SAT scores that are 5.54 points [95\% CI 8.82 to 2.26] lower.  

But the real story is hidden behind one of the ``other factors" that we warn students about but don't generally teach how to address!  The proportion of students taking the SAT varies dramatically by region, as do teacher salaries.  
In the midwest and plains states, where teacher salaries tend to be lower, relatively few high school students take the SAT exam.  These are 
typically the top students who are planning to attend college out of state, while many others take the 
alternative standardized ACT test.  The net result is that the fraction
taking the SAT is a confounding factor.  In a multiple regression model that controls for this variable, the
sign of the slope parameter flips.  The new model predicts that a state with an average salary that is one 
thousand dollars higher than another would expect to have SAT scores that are 2.18 points higher [95\% CI 0.11 to  4.25].

This problem is a continuous example of Simpson's paradox.
There is no automated data mining or machine learning methods that can get close to the right answer if the stratifying variable is not collected. However, statistical thinking with an appreciation of Simpson's paradox would alert a student to \emph{look for} the hidden confounding variables.
To tackle this problem, students need to understand multivariable modeling.

One natural approach to develop such understanding is multiple regression.
While this is not a traditional topic included in introductory statistics, an increasing number of textbooks
and courses are incorporating the basic principles (often purely as a descriptive summarization of the data).  

Another option for this analysis is even simpler: the use of stratification. 
We can arbitrarily split the states up into groups based on the fraction of
students taking the SAT.  The rightmost scatterplot in Figure \ref{fig:sat} displays a grouping of 
states with 0-22\% of students (``low fraction", top line), 23-49\% of students (``medium fraction", middle line), and
50-81\% (``high fraction", bottom line). The story is clear: there is a positive or flat relationship between teacher salary
and SAT score for each of these groups, but when we average over them, we observe a negative relationship.

Shattuck would have recognized this problem: the mortality estimates in Figure \ref{fig:shattuck} 
are multivariate, with discrete strata of sub-populations.  
This type of multivariable thinking is critical to make sense of the observational data around us.  If students
don't see some tools for disentangling complex relationships, they may dismiss statistics as an old-school
discipline only suitable for small sample inference of randomized studies.
Without a background in these key design topics, 
data scientists may make errors in their interpretations.

\subsection{Data-related skills}

Statistics is a science without a practicum: other scientists work with pipettes and learn other practical skills to allow them to apply their
theories to the real world.  
Statisticians need to be facile with computation to develop and refine the practical skills that allow them to apply theory to the real world.
Increasingly, students need experience analyzing larger, real-world data sets and to be aware of what techniques do and do not scale well.  

To be effective, students need to develop data habits of mind \cite{finz:2013}.  They need to be able to
think creatively about data and understand conceptions of ``data tidying" \cite{wick:2014}.  
When working with data, students must first determine the question, describe a solution in terms that a computer can understand, and execute the commands to implement the solution \cite{asa:2014}.
They need
facility with data sets of varying sizes along with experiences wrestling with large, messy, complex, and
challenging datasets \cite{hort:baum:2015}.

The statistical data analysis cycle is an iterative process.  It involves the formulation of questions, data collection, data cleaning and derivation, exploratory analysis, modeling, and interpretation and communication of results. 
Hadley \citeasnoun{wick:2015} has argued that analysts need to be able to:
\begin{description}
\item[ingest:] load data,
\item[manipulate:] filter, summarize, etc.,
\item[visualize:] explore data,
\item[model:] answer a precise question with a model, and
\item[report:] communicate the solution to others.
\end{description}

Visualization of data is important not only for final reports, but as an integral part of the exploration, error-checking, and analysis cycle \cite{nola:2015}.  During the analysis phase, visualization and modeling 
form an important pair.
Our students need both technical (computational) visualization skills and general visualization strategies that are transferable from one technological tool to another but are not necessarily developed just by using a visualization tool.

These changes may force instructors out of their comfort zones as they need to become familiar with databases, XML, web scraping, and other data technologies that go beyond simple manual curation of spreadsheets.  This is not rocket science: recent developments have dramatically decreased the degree of difficulty involved in such technical topics \cite{hort:baum:2015}.
Instructors and programs will need to stay up to date as the preferred technologies change over time.  
Instructors and institutions that provide these skills will get the students.
Students with these skills will get the jobs.  

We need to introduce these capacities in our first and second courses to ensure that students have sufficient depth by the end of their programs.
Integrative capstone experiences are an excellent 
place to integrate (but not to introduce!) these skills, see \citeasnoun{laza:2011}.  In addition, co-curricular events can help
to provide opportunities for our students to refine their ability to ``wrangle" data.
The ASA DataFest weekend-long ``celebrations of data" (\url{http://www.amstat.org/education/datafest}), founded by Rob Gould and colleagues
at UCLA, provide opportunities for undergraduate students to provide meaning for data and motivate study of these
data-related skills.

\subsection{Empirical simulation to complement analytic problem solving}

Statistics as a discipline has expanded far beyond what any one person can individually master.  Our graduate
and undergraduate programs (as well as introductory and intermediate courses) must provide
a framework that can be used to learn new topics, methods, and approaches.  So how do we train students to be 
able to be life-long learners?  
Besides providing a useful check on analytic answers, simulations may
provide insights into how to solve a problem \cite{hort:2013}.
I believe that use of empirical problem solving, using computational tools 
and simulation, may free up aspects of our curriculum and allow students to be nimbler and better able to find answers to problems that didn't exist when they were trained.

Consider an example from the excellent probability and mathematical statistics text by John \citeasnoun{rice:2006}.
I've repeatedly adopted this book, plan to do so in the future, and continue to highly recommend it.  But one exercise is highly illustrative of the challenges
and opportunities of what and how we teach.

(Problem 3.11) Let A, B, and C be independent random variables each distributed uniform in the interval [0,1].  
Question: What is the probability that the roots of the quadratic equation given by $Ax^2 + Bx + C = 0$ are real? 

We can begin with the empirical solution.  
This is very straightforward to simulate in R (or similar environment), after noting that the roots will only
be real if the discriminant $B^2 - 4AC$ in the quadratic formula is non-negative.
\begin{knitrout}
\definecolor{shadecolor}{rgb}{0.969, 0.969, 0.969}\color{fgcolor}\begin{kframe}
\begin{alltt}
\hlstd{numsim} \hlkwb{<-} \hlnum{1000000}
\hlstd{A} \hlkwb{<-} \hlkwd{runif}\hlstd{(numsim); B} \hlkwb{<-} \hlkwd{runif}\hlstd{(numsim); C} \hlkwb{<-} \hlkwd{runif}\hlstd{(numsim)}

\hlstd{discrim} \hlkwb{<-} \hlstd{B}\hlopt{^}\hlnum{2} \hlopt{-} \hlnum{4}\hlopt{*}\hlstd{A}\hlopt{*}\hlstd{C}
\hlstd{realroot} \hlkwb{<-} \hlstd{discrim} \hlopt{>=} \hlnum{0}
\hlkwd{table}\hlstd{(realroot)}\hlopt{/}\hlstd{numsim}
\end{alltt}
\begin{verbatim}
## realroot
##    FALSE     TRUE 
## 0.745858 0.254142
\end{verbatim}
\end{kframe}
\end{knitrout}

Not surprisingly, when we run the simulation again, we get a (slightly) 
different answer (in this case 0.254227).
The true value appears to be somewhere in the range of 0.254.

Next we consider one analytic solution.
We begin by defining $Y = B^2$ and $W = 4AC$.  The distribution of Y is given by:
$$f(y) =
\left\{\begin{array}{ll }
\frac{1}{2\sqrt{y}} & \mathrm{if} \ 0 \leq y \leq 1 \\
0 & \mathrm{otherwise}\\
\end{array}
\right.$$
The distribution of W is given by:
$$ f(w) =
\left\{ \begin{array}{ll }
-\log{(w/4)/4} & \mathrm{if} \ 0 \leq w \leq 4 \\
0 & \mathrm{otherwise} \\
\end{array}
\right.$$
Since $Y$ and $W$ are independent, the joint distribution is given by:
$$ f(y, w) =
\left\{ \begin{array}{ll }
\frac{-\log{(w/4)}}{8 \sqrt{y}} & \mathrm{if} \ 0 \leq y \leq 1  \ \mathrm{and} \ 0 \leq w \leq 4\\
0 & \mathrm{otherwise} \\
\end{array}
\right.$$
The discriminant $B^2 - 4AC$ is non-negative when $Y > W$.
\begin{eqnarray*}
P(Y > W) &=& \int_0^1 \int_0^y f(y, w) \ dw \, dy \\
&=& \int_0^1 \int_0^y \frac{-\log{(w/4)}}{8 \sqrt{y}} \ dw \, dy \\
&=& \int_0^1 \frac{\sqrt{y}(-\log{(y)} + 1 + \log{(4)})}{8} \, dy \\
&=& \frac{5+\log{(64)}}{36} \approx 0.254413.
\end{eqnarray*}

In a perfect world, students would be able to tackle problems both ways (though one might argue that the computational
solution is far simpler and requires far fewer mathematical prerequisites).  

What's instructive for this example is the fact that the answer in the back of the first, second, and third
editions of Rice's otherwise superb book is given as 1/9.  While I suspect that this was a transcription error as 
the solutions were compiled, it's illustrative that over multiple editions the incorrect answer was provided as the solution of
a problem from the third chapter
of a widely adopted text (note: the correct answer is now provided in the online errata).

What are the implications and ramifications of this motivating exercise?  I see several:
\begin{enumerate}
\item It's hard to get problems of this sort wrong if you check them using simulations,
\item Many phenomena (including a number that are not analytically tractable) are amenable to simulation, 
\item While it's still important to be able to get the exact answer (and not just an approximation), it may not be as necessary for all problems (particularly at the undergraduate level), 
\item For many models that involve stochastic processes, the choice is often between simulation answers from comparatively realistic models or analytic answers from oversimplified models, and
\item Instructors should work to develop parallel empirical and analytical problem-solving skills \cite{hort:brow:2004,hort:2013}.
\end{enumerate}

Many areas of modern statistics (e.g. resampling based tests,
Bayesian inference, model diagnostics and assessment, etc.) 
can be explored
by students with only a modicum of programming skills \cite{hort:2013}.
Others have made similar arguments: \citeasnoun{cars:hard:2015} describe how such simulation-based approaches
can be valuable for political science graduate students.
\citeasnoun{nola:spee:1999} and \citeasnoun{nola:spee:2000} proposed a model of extended case studies
to encourage and develop statistical thinking in combination with computation.
The Math Sciences 2025 report \cite{mathsci2025} called for more mathematical scientists with experience with
computation and noted that ``the ability to simulate a phenomenon is often regarded as a test of our ability to understand it" (p.\,74).  Moving from a small portion of data-related and computational skills to a 
full platter will help the use of statistics flourish.

\section{Closing thoughts}

Getting students to come to grips with multidimensional thinking, preparing them to grapple with real-world problems
and complex data, and providing them with skills in computation are challenging things to add to our curriculum.
But such an approach would help them to tackle more sophisticated problems, assess their models and assumptions, carry out sensitivity analyses, and check their results.
In addition, students need to 
develop the capacity to work effectively in groups and communicate their results \cite{asa:2014}.  
If we are able to restructure
and reformulate our curriculum (without losing the key components that define us as a profession),
we will be an integral part of the expanding world of data science.  Statisticians have a lot to offer to data science, particularly with respect to making it more rigorous, scientific, and reproducible.

But how do we make this happen?  What structures are in place to help facilitate the necessary changes to 
curricula, train new teachers, and oversee these efforts?
The ASA
and its members play a key role in fostering such changes.  

At our founding, the ASA was a home for some odd ducks who were actively developing methods that are still bearing fruit.  
One hundred years later in 1939, the association and the profession were engaged with questions that still resonate today: development of statistics as a field of study, professional approaches available but not often used, and challenges of computation.

Now we face the challenges of `Big Data' and the need for data-related skills.
In his video introduction to the keynote for the Strata+Hadoop Big Data Conference in 2015, 
President Barack Obama stated that ``understanding and innovating with data has the potential to change how we do almost anything for the better."
I heartily concur.  But the word ``statistics" was not mentioned in his presentation.  

Statistics and statistics education have been evolving since our founding.  New opportunities and new threats are emerging, which necessitate more evolution.  We need to bring the approaches refined over the history of our discipline to bear to address these challenges.

In the second part of ``Democracy in America" (published around the time of the ASA's founding in 1839), Alexis de Tocqueville shared observations about public associations in the United States:
\begin{quote}
The political associations which exist in the United States are only a
single feature in the midst of the immense assemblage of associations in that
country; Americans of all ages, all conditions, and all dispositions, constantly
form associations. (p. 129)

They have not only commercial and manufacturing companies, in which
all take part, but associations of a thousand other kinds, religious, moral,
aerious, futile, general or restricted, enormous or diminutive. (p. 129) 

Feelings and opinions are recruited, the heart is enlarged, and the human
mind is developed, only by the reciprocal influence of men upon each other.
(p.\,132)

As soon as several of the inhabitants of the United States have taken up
an opinion or a feeling which they wish to promote in the world, they look
out for mutual assistance; and as soon as they have found each other out,
they combine. (p. 133)
\cite{deto:1840,davi:1940}
\end{quote}

The ASA serves as an important public organization to help to guide us, as individuals and institutions, towards a shared sense for where things are heading.  It also provides the professional support to assist as we change and adapt.

Much has changed since the time of our founding.  Women and men are now fully engaged in the statistics profession
(with 45\% of our undergraduate degrees awarded to women (see \url{http://www.amstat.org/newsroom/pressreleases/2015-StatsFastestGrowingSTEMDegree.pdf}), new sources of 
data allow us to make better decisions, and 
metaphors of ``big tents" \cite{rodr:2013} rather than ``enlarged hearts" are at the center of our discourse.
There are important challenges and opportunities before us, but I feel confident that
we are in a position to address them.
The value of shared connections fostered by an association still has tremendous value.
This is an exciting time to be a statistician, and I look forward to what will transpire in the coming years.

\section*{Acknowledgements}

This work was supported by NSF grant 0920350 (Phase II: Building a Community around Modeling, Statistics, Computation, and Calculus).  
Thanks to
Andrew Bray,
George Cobb, 
Kay Endriss,
Joan Garfield,
Garrett Grolemund,
Johanna Hardin,
Danny Kaplan, 
John McKenzie,
Xiao-Li Meng,
Marcello Pagano,
Randall Pruim,
Jean Riseman, 
Nathaniel Schenker, 
Jessica Utts, 
Chris Wild,
Jeffrey Witmer,
Andrew Zieffler, and the editor
for suggestions and helpful comments 
on an earlier draft.  

\singlespacing
\bibliographystyle{dcu}
\bibliography{hthesis}

\end{document}